\documentclass[aps,pr,twocolumn,psfig,eqsecnum,epsf]{revtex4}
\usepackage{graphicx}
\usepackage{color}

\usepackage[version=3]{mhchem} 
\usepackage{amssymb}

\begin{document}
\title{\Large \bf Electric double layer for spherical particles in salt-free concentrated suspensions including ion size effects}
\author{ \large Rafael Roa,$^1$ F\'elix Carrique,$^{1,}$\footnote{E-mail address: carrique@uma.es} and Emilio Ruiz-Reina$^2$\\} 
\affiliation{ $^1$ F\'{\i}sica Aplicada I, Universidad de M\'alaga, Spain,\\
$^2$ F\'{\i}sica Aplicada II, Universidad de M\'alaga, Spain.}
\date{\today}

\vspace{0.5cm} 
\begin{abstract}                
The equilibrium electric double layer (EDL) that surrounds the colloidal particles is determinant for the response of a suspension under a variety of static or alternating external fields. An ideal salt-free suspension is composed by the charged colloidal particles and the ionic countercharge released by the charging mechanism. The existing macroscopic theoretical models can be improved by incorporating different ionic effects usually neglected in previous mean-field approaches, which are based on the Poisson-Boltzmann equation (PB). The influence of the finite size of the ions seems to be quite promising because it has been shown to predict phenomena like charge reversal, which has been out of the scope of classical PB approximations. In this work we numerically obtain the surface electric potential and the counterions concentration profiles around a charged particle in a concentrated salt-free suspension corrected by the finite size of the counterions. The results show the large importance of such corrections for moderate to high particle charges at every particle volume fraction, specially, when a region of closest approach of the counterions to the particle surface is considered. We conclude that finite ion size considerations are obeyed for the development of new theoretical models to study nonequilibrium properties in concentrated colloidal suspensions, particularly the salt-free ones with small and highly charged particles.
\end{abstract} 
\maketitle
\vspace{0.5cm}


\section{Introduction}
It is well-known that many transport phenomena in suspensions of charged particles, particularly their stability and electrokinetic response, are closely related to the properties of the equilibrium electric double layer (EDL) surrounding the particles \cite{Lyklema1995,Hunter1995,Masliyah2006}. Most of these theoretical EDL models are based on the classical Poisson-Boltzmann equation (PB), a mean-field approach which has been the cornerstone for the study of the diffuse part of the EDL near charged interfaces in electrolyte solutions.

Many of the studies concern suspensions with low particle concentration, but nowadays it is the concentrated regime \cite{OBrien1990,Ohshima2006,Dukhin1999,Carrique2003} what deserves more attention because of its practical applications. From the theoretical point of view, these systems lack of a full understanding due to the inherent complexity associated with the increasing particle-particle electrohydrodynamic interactions as particle concentration grows.

Very commonly, the electrical interaction between particles is largely screened at typical electrolyte concentrations, and particle ordering is impeded. However, if the concentration of external salt is sufficiently low, or there is not any salt added to the suspension, the electrical interactions can give rise to different short or long-ranged ordered phases at relatively low particle volume fractions. These systems are usually called colloidal crystals or glasses \cite{Sood1991,Medebach2003}. Suspensions composed just by the charged particles and their ionic countercharge (the so-called Òadded counterionsÓ) in the liquid medium are named salt-free suspensions, and its study has increasingly grown in recent years from experimental and theoretical points of view.

The theoretical models for such systems are generally scarce in comparison with those for suspensions in electrolyte solutions. In these systems there is a coupling between the added counterions released by each particle and the opposite charge on the particle surface to preserve the electroneutrality. If the suspension is concentrated \cite{Ohshima2006}, the particle-particle electrohydrodynamic interactions introduce severe limitations when dealing with the response to external fields, not only for the mathematical difficulties of dealing with many-body interactions, but also for the additional numerical problems raised when it comes to solve the equations, no matter the approximation chosen. The authors have addressed this problem by using a cell model approach to deal with electrokinetic and rheological properties in concentrated suspensions \cite{Carrique2006,Carrique2008,Carrique2008b,RuizReina2007,RuizReina2008,Carrique2009,Carrique2009b,RuizReina2010,Carrique2010}.

In spite of the achievements of the cell model, it maintains the drawbacks of the PB treatment, particularly, the neglecting of ionic correlations and the consideration of point-like ions in solution. Fortunately, for many situations including low to moderate particle charges and monovalent ions in solution, this approach provides an excellent representation of the equilibrium problem. On the contrary, it has been argued that it yields unphysical ionic concentration profiles near highly charged interfaces, and also that it is not able to predict important phenomena like, for example, the overcharging, also called charge reversal or charge inversion \cite{IbarraArmenta2009}. In the literature we can find different attempts to overcome these PB limitations. Some of them concerns microscopic descriptions of ion-ion correlations and the finite size of the ions \cite{IbarraArmenta2009,Chodanowski2001,JimenezAngeles2004,Diehl2008,MartinMolina2009}. Others are based on macroscopic descriptions considering average interactions by mean-field approximations, that include entropic contributions related with the excluded volume effect when the ions have a finite size \cite{Bikerman1942,Outhwaite1983,Adamczyk1996,Borukhov1997,Borukhov2000,Borukhov2004,Lue1999,KraljIglic1996,Bohinc2001,Biesheuvel2007,Bazant2009,LopezGarcia2007,LopezGarcia2008,ArandaRascon2009,ArandaRascon2009b,Bhuiyan2004,Bhuiyan2009}. While the former ones are physically more robust to describe real systems, they are basically restricted to equilibrium conditions. The macroscopic approaches, however, permit us to make predictions under equilibrium and non-equilibrium conditions. Unfortunately, it seems that mean-field approaches eliminate some of the crucial information necessary for explaining, for example, the overcharging phenomena. In the past, it has been shown with Monte Carlo simulations, and also with some theoretical approximations like the HNC-MSA \cite{GonzalezTovar1989}, that electrical ion-ion correlations joined to exclude ion volume effects could be the key point to understand such phenomena \cite{IbarraArmenta2009}.

In a series of papers, L\'opez-Garc\'ia and coworkers \cite{LopezGarcia2007,LopezGarcia2008,ArandaRascon2009,ArandaRascon2009b}, studied the equilibrium EDL of a particle in a dilute colloidal suspension in an electrolyte solution accounting for the finite size of the ions by using a modified Poisson-Boltzmann equation (MPB). They also assumed the existence of a common distance of closest approach to the particle surface for all ionic species. Their first results indicated that the associated exclude volume effect was very important for equilibrium and non-equilibrium properties. Shortly after, their model was compared with some Monte Carlo simulations for general ionic conditions \cite{IbarraArmenta2009}, and while the latter predicted charge reversal for the case of multivalent ions, the MPB was unable to show such behaviours whatever the valence of the ions may be. It was concluded that the neglecting of ion-ion correlations in the MPB model was its main drawback.

Quite recently, L\'opez-Garc\'ia \textit{et al.} \cite{LopezGarcia2010} have extended their previous model to allow for a different distance of closest approach to the particle surface for each ionic species. The most important fact is that it predicts charge reversal under appropriate circumstances by only considering such ionic excluded volume effects. To our knowledge, it is the first time that a phenomenological theory based on macroscopic descriptions for equilibrium and non-equilibrium properties in colloidal suspensions is able to predict charge inversion, mainly with multivalent ions. In other words, the charge reversal phenomenon, that has been elusive for such formalisms, now displays a reasonable explanation from the macroscopic approach based on mean-field approximations. No consideration of explicit ion-ion correlations is then required. Although no charge reversal is possible in an ideal salt-free suspension because the whole ionic countercharge equals that of the particle, and cannot exceed it, these achievements support our choice of using a similar macroscopic approach for our salt-free concentrated suspensions, with the help of a cell model. It has been recently shown that, up to moderately strong electrostatic couplings, the cell model proves accurate for the prediction of osmotic pressures of deionised suspensions in agreement with Monte Carlo simulations and renormalized-effective interaction approaches \cite{Denton2010}. 

The macroscopic approach provides reasonable predictions for many non-equilibrium phenomena in concentrated suspensions, which are in good agreement with experiments. Other descriptions, like Monte Carlo's or microscopic models, are unable of making feasible  predictions out of equilibrium. Summarizing, in this work we aim at studying the EDL of a particle in a concentrated salt-free suspension considering that the added counterions have finite ion size. Our approach is a mean-field one with an adequate MPB equation inside the cell. The procedure is a generalization of that already used by Borukhov \cite{Borukhov2004} for the special case of a salt-free suspension, valid for the concentrated case. Unlike Borukhov's treatment, our model also incorporates an excluded region in contact with the particle of a hydrated radius size, which has been shown by Aranda-Rasc\'on \textit{et al.} \cite{ArandaRascon2009,ArandaRascon2009b} to yield a more realistic representation of the solid-liquid interface and also to predict results in better agreement with experimental electrokinetic data. The equations will be numerically solved and the equilibrium surface potential and counterion concentration profiles inside a cell will be analyzed upon changing particle volume fraction, particle surface charge density, and size of the counterions. In order to show the realm of the finite ion size effect, the results will be compared with MPB predictions that do not take into account a finite distance of closest approach, and also with standard PB predictions for point-like ions.

\section{Theory}
\subsection{The cell model}
To account for the interactions between particles in concentrated suspensions, a cell model is used (bare Coulomb interactions among particles are included in an average sense, but ions-induced interactions between particles as well as ion-ion correlations, are ignored). For details about the cell model approach see the excellent review of Zholkovskij \textit{et al.} \cite{Zholkovskij2007}. Different electrokinetic and rheological phenomena in salt-free concentrated suspensions have been developed by the authors according to this concept: the DC electrokinetic response to static electric fields \cite{Carrique2006} (electrophoretic mobility and electrical conductivity), their response to oscillating electric fields \cite{Carrique2008,Carrique2008b} (dynamic mobility, complex conductivity and dielectric response); the electroviscous effect \cite{RuizReina2007}, and some related to more realistic descriptions of the phenomena taking into account the possibility of chemical reactions in the system in agreement with experimental conditions \cite{RuizReina2008,Carrique2009,Carrique2009b,RuizReina2010,Carrique2010}. We have learned from those works how important the description of the EDL is for the non-equilibrium responses. 

\begin{figure}[t]
\centering
  \includegraphics[height=4.5cm]{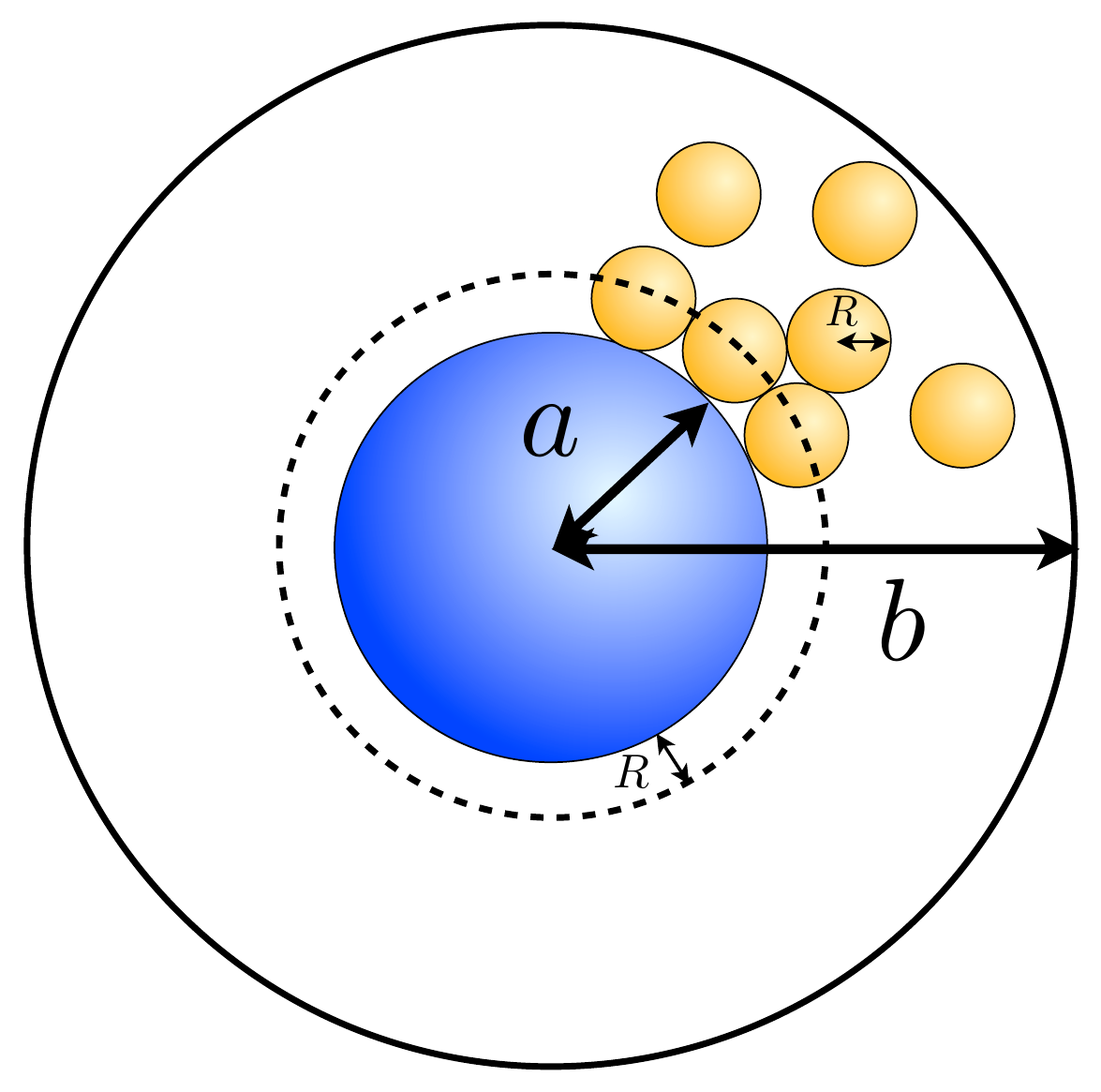}
  \caption{Cell model including the distance of closest approach of the counterions to the particle surface.}
  \label{fgr:cell}
\end{figure}

Concerning the  cell model, Fig. \ref{fgr:cell}, each spherical particle of radius $a$ is surrounded by a concentric shell of the liquid medium, having an outer radius $b$ such that the particle/cell volume ratio in the cell is equal to the particle volume fraction throughout the entire suspension, that is

\begin{equation}\label{phidef}
\phi=\left( \frac{a}{b}\right)^3
\end{equation}

The basic assumption of the cell model is that the macroscopic properties of a suspension can be obtained  from appropriate averages of local properties in a unique cell.

\subsection{Finite size of the counterions}
Let us consider a spherical charged particle of radius $a$ and surface charge density $\sigma$ immersed in a salt-free medium with only the presence of the added counterions of valence $z_c$. The axes of the spherical coordinate system ($r$, $\theta$, $\varphi$) are fixed at the center of the particle. In the absence of any external field, the particle is surrounded by a spherically symmetrical charge distribution.

Within a mean-field approximation, the total free energy of the system, $F=U-TS$, can be written in terms of the equilibrium electric potential $\Psi(\mathbf{r})$ and the counterion concentration $n_c(\mathbf{r})$. The configurational internal energy contribution $U$ is
\begin{equation}
  U=\int \mathrm{d}\textbf{r}\bigg{[}-\frac{\epsilon_0\epsilon_r}{2}|\nabla\Psi(\mathbf{r})|^2+z_cen_c(\mathbf{r})\Psi(\mathbf{r})-\mu_cn_c(\mathbf{r})\bigg{]}
\end{equation}

The first term is the self-energy of the electric field, where $\epsilon_0$ is the vacuum permittivity, and $\epsilon_r$ is the relative permittivity of the suspending medium. The next term is the electrostatic energy of the counterions in the electrostatic mean field, and the last term couples the system to a bulk reservoir, where $\mu_c$ is the chemical potential of the counterions.

The entropic contribution $-TS$ is
\begin{multline}
  -TS=k_BTn_c^{max}\int \mathrm{d}\textbf{r}\bigg{[}\frac{n_c(\mathbf{r})}{n_c^{max}}\ln \left(\frac{n_c(\mathbf{r})}{n_c^{max}}\right)\\ 
  +\left(1-\frac{n_c(\mathbf{r})}{n_c^{max}}\right)\ln \left(1-\frac{n_c(\mathbf{r})}{n_c^{max}}\right)\bigg{]}
\end{multline}
where $k_B$ is Boltzmann's constant, $T$  is the absolute temperature, and $n_c^{max}$ is the maximum possible concentration of counterions due to the excluded volume effect, defined as $n_c^{max}=V^{-1}$, where $V$ is the average volume occupied by an ion in the solution. The first term inside the integral is the entropy of the counterions, and the second one is the entropy of the solvent molecules. This last term accounts for the ion size effect that modifies the classical Poisson-Boltzmann equation and was proposed earlier by Borukhov \textit{et al.} \cite{Borukhov1997}

The variation of the free energy $F=U-TS$ with respect to $\Psi(\mathbf{r})$ provides the Poisson equation
\begin{equation}\label{poisson}
  \nabla^2\Psi(\mathbf{r})=-\frac{z_ce}{\epsilon_0\epsilon_r}n_c(\mathbf{r})
\end{equation}
and the counterions concentration is obtained performing the variation of the free energy with respect to $n_c(\mathbf{r})$
\begin{equation}\label{nc}
  n_c(\mathbf{r})=\frac{b_c\exp\left(-\frac{z_ce\Psi(\mathbf{r})}{k_BT}\right)}{1+\frac{b_c}{n_c^{max}}\left[\exp\left(-\frac{z_ce\Psi(\mathbf{r})}{k_BT}\right)-1\right]}
\end{equation}
where $b_c$ is an unknown coefficient that represents the ionic concentration where the electric potential is zero.

Applying the spherical symmetry of the problem and combining eqns (\ref{poisson}) and (\ref{nc}), we obtain the modified Poisson-Boltzmann equation (MPB) for the equilibrium electric potential
\begin{multline}\label{mpb}
\frac{\mathrm{d}^2\Psi(r)}{\mathrm{d}r^2}+\frac{2}{r}\frac{\mathrm{d}\Psi(r)}{\mathrm{d}r}\\ =-\frac{z_ce}{\epsilon_0\epsilon_r}\frac{b_c\exp\left(-\frac{z_ce\Psi(r)}{k_BT}\right)}{1+\frac{b_c}{n_c^{max}}\left[\exp\left(-\frac{z_ce\Psi(r)}{k_BT}\right)-1\right]}
\end{multline}

As the cell is electrically neutral we have
\begin{align}
Q&=4\pi a^2\sigma=-4\pi \int_a^b \rho (r)r^2\mathrm{d}r \nonumber \\
&=-4\pi z_ce \int_a^b n_c (r)r^2\mathrm{d}r 
\end{align}

The unknown coefficient $b_c$ can be calculated iteratively substituting the value of $n_c (r)$, eqn (\ref{nc}), in the last expression  
\begin{equation}\label{electroneutrality}
\sigma=-\frac{z_ce}{a^2}\int_a^b\frac{b_c\exp\left(-\frac{z_ce\Psi(r)}{k_BT}\right)}{1+\frac{b_c}{n_c^{max}}\left[\exp\left(-\frac{z_ce\Psi(r)}{k_BT}\right)-1\right]}r^2\mathrm{d}r
\end{equation}

We need two boundary conditions to solve the MPB equation. The first one is
\begin{equation}\label{dpot-b}
\frac{\mathrm{d}\Psi(r)}{\mathrm{d}r}\bigg |_{r=b}=0
\end{equation}
which derives from the electroneutrality condition of the cell and the application of Gauss theorem to the outer surface of the cell. The second one is
\begin{equation}\label{pot-b}
\Psi(b)=0
\end{equation}
that fixes the origin of the electric potential at $r=b$. 

The MPB problem, eqns (\ref{mpb}), (\ref{dpot-b}) and (\ref{pot-b}), can be solved iteratively using eqn (\ref{electroneutrality}) to find the unknown coefficient $b_c$. Fortunately, the authors have recently shown \cite{RuizReina2008} that this kind of problem can be solved avoiding the iterative procedure. For this purpose, we will use dimensionless variables, which are defined as
\begin{eqnarray}\label{adim}
x=\frac{r}{a} \qquad \tilde{\Psi}(x)=\frac{e\Psi(r)}{k_BT} \qquad \tilde{\sigma}=\frac{ea}{\epsilon_0\epsilon_rk_BT}\sigma  \nonumber \\
\tilde{b}_c=\frac{e^2a^2}{\epsilon_0\epsilon_rk_BT}b_c \qquad \tilde{n}_c^{max}=\frac{e^2a^2}{\epsilon_0\epsilon_rk_BT}n_c^{max}
\end{eqnarray}
rewritting eqn (\ref{mpb}) as
\begin{equation}\label{mpbad}
g(x)\equiv\frac{\mathrm{d}^2\tilde{\Psi}(x)}{\mathrm{d}x^2}+\frac{2}{x}\frac{\mathrm{d}\tilde{\Psi}(x)}{\mathrm{d}x}=\frac{-z_c\tilde{b}_c\mathrm{e}^{-z_c\tilde{\Psi}(x)}}{1+\frac{\tilde{b}_c}{\tilde{n}_c^{max}}\left(\mathrm{e}^{-z_c\tilde{\Psi}(x)}-1\right)}
\end{equation}
where we have defined the function $g(x)$. If we differentiate it, after a little algebra, it is possible to eliminate the unknown coefficient $\tilde{b}_c$ and find that 
\begin{equation}\label{gprime}
g'(x)+z_cg(x)\tilde{\Psi}'(x)+\frac{1}{\tilde{n}_c^{max}}g^2(x)\tilde{\Psi}'(x)=0
\end{equation}
where the prime stands for differentiation with respect to $x$. In terms of the electric potential, eqn (\ref{gprime}) is rewritten as
\begin{multline}\label{mpb3}
\tilde{\Psi}'''(x)+\frac{2}{x}\tilde{\Psi}''(x)-\frac{2}{x^2}\tilde{\Psi}'(x)\\ 
+\tilde{\Psi}'(x)\left(\tilde{\Psi}''(x)+\frac{2}{x}\tilde{\Psi}'(x)\right)\\ 
\cdot\left[z_c+\frac{1}{\tilde{n}_c^{max}}\left(\tilde{\Psi}''(x)+\frac{2}{x}\tilde{\Psi}'(x)\right)\right]=0
\end{multline}
Eqn (\ref{mpb3}) is a nonlinear third-order differential equation that needs three boundary conditions to completely specify the solution. Two of them are provided by eqns (\ref{dpot-b}) and (\ref{pot-b}), which now read
\begin{equation}\label{bcmpb31}
\tilde{\Psi}'(h)=0 \qquad \tilde{\Psi}(h)=0 
\end{equation}
where $h=(b/a)=\phi^{-1/3}$ is the dimensionless outer radius of the cell. The third one specifies the electrical state of the particle, and can be obtained by applying Gauss theorem to the outer side of the particle surface $r=a$ 
\begin{equation}\label{dpot-a}
\frac{\mathrm{d}\Psi(r)}{\mathrm{d}r}\bigg |_{r=a}=-\frac{\sigma}{\epsilon_0\epsilon_r}
\end{equation}

Its dimensionless form is
\begin{equation}\label{bcmpb32}
 \tilde{\Psi}'(1)=-\tilde{\sigma}
\end{equation}

Regarding dilute suspensions, it is very common in the literature to use the surface charge (also regulated surface charge) or the surface potential as a boundary condition at the particle surface when solving the equilibrium Poisson-Boltzmann equation, and all of them are equally valid. When it comes to concentrated suspensions, we prefer the use of the particle surface charge as boundary condition by the following reasoning. The particle charge is a property that can be often measured experimentally. Instead, the surface potential depends on the choice of the potential origin, which is the solution ``bulk'' for the dilute case (see the discussion in the Appendix). As in the concentrated case there is not a clear ``bulk'', the origin is typically chosen at the outer surface of the cell, $r=b$, which depends on the particle volume fraction, eqn (\ref{phidef}). In any case, we always can use the surface potential as a boundary condition, representing the potential difference between the particle surface and the outer surface of the cell.

If we consider point-like counterions, $n_c^{max}=\infty$, eqn (\ref{mpb3}) becomes the expression obtained by Ruiz-Reina and Carrique \cite{RuizReina2008}. It can be easily demonstrated that solving the third-order problem is mathematically equivalent to finding the solution of the iterative problem, that is, that any function satisfying eqns (\ref{mpb3}), (\ref{bcmpb31}) and (\ref{bcmpb32}) also satisfies eqns (\ref{mpb}) and (\ref{electroneutrality})$-$(\ref{pot-b}), and vice versa. We have removed the unknown coefficient $b_c$ using this procedure and now the electric potential can be obtained numerically in one single step. 

Eqns (\ref{mpb3}), (\ref{bcmpb31}) and (\ref{bcmpb32}) form a boundary value problem that can be solved numerically using the MATLAB routine bvp4c \cite{Kierzenka2001}, that computes the solution with a finite difference method by the three-stage Lobatto IIIA formula. This is a collocation method that provides a $C^1$ solution that is fourth order uniformly accurate at all the mesh points. The resulting mesh is non-uniformly spaced out and has been chosen to fulfill the admitted error tolerance (taken as $10^{-6}$ for all the calculations).

Once we have found the electric potential $\tilde{\Psi}(x)$ we can obtain the coefficient $\tilde{b}_c$ by evaluating eqn (\ref{mpbad}) at $x=h$
\begin{equation} \label{bc}
\tilde{b}_c=-\frac{\tilde{\Psi}''(h)}{z_c}
\end{equation}

The ionic concentration $n_c(r)$ is determined using eqns (\ref{nc}) and (\ref{adim}).

\subsection{Excluded region in contact with the particle}
Following the work of Aranda-Rasc\'on \textit{et al.} \cite{ArandaRascon2009}, we incorporate a distance of closest approach of the counterions to the particle surface, resulting from their finite size. We assume that counterions cannot come closer to the surface of the particle than their effective hydration radius, $R$, and, therefore, the ionic concentration will be zero in the region between the particle surface, $r=a$, and the spherical surface, $r=a+R$, defined by the counterion effective radius. This reasoning implies that counterions are considered as spheres of radius $R$ with a point charge at its center.

The whole electric potential $\Psi(r)$ is now determined by the stepwise equation
\begin{equation}\label{laplace2}
\begin{cases}
\frac{\mathrm{d}^2\Psi(r)}{\mathrm{d}r^2}+\frac{2}{r}\frac{\mathrm{d}\Psi(r)}{\mathrm{d}r}=0 & \text{ \ } a\leq r \leq a+R \\
& \\
\text{Eqn }(\ref{mpb}) & \text{ \ } a+R\leq r \leq b
\end{cases}
\end{equation}

To completely specify the problem we now should force the potential to be continuous at the surface $r=a+R$, and also its first derivative, which is related to the continuity of the normal component of the electric displacement at that surface, in addition to boundary conditions, eqns (\ref{dpot-b}) and (\ref{pot-b}). Thus, in the region in contact with the particle $[a,\ a+R]$, we are solving Laplace's equation, and, in the region $[a+R,\ b]$, the MPB equation that we obtained previously in eqn (\ref{mpb}).

As we have seen before, we can eliminate the coefficient $b_c$ changing the system of second order differential equations, eqn (\ref{laplace2}), into one of third order, which hugely simplify the resolution process
\begin{equation}\label{laplace3}
\begin{cases}
\tilde{\Psi}'''(x)+\frac{2}{x}\tilde{\Psi}''(x)-\frac{2}{x^2}\tilde{\Psi}'(x)=0 & \text{ \ } 1\leq x \leq 1+\frac{R}{a} \\
& \\
\text{Eqn }(\ref{mpb3}) & \text{ \ } 1+\frac{R}{a}\leq x \leq h
\end{cases}
\end{equation}
where we have use dimensionless variables. The boundary conditions needed to completely close the problem are
\begin{eqnarray}\label{bclaplace3}
\tilde{\Psi}'_L(1)=-\tilde{\sigma} && \tilde{\Psi}'_L(1+\tfrac{R}{a})=\frac{-\tilde{\sigma}}{(1+R/a)^2} \qquad \nonumber\\ 
\tilde{\Psi}_{P}(h)=0 && \tilde{\Psi}'_{P}(h)=0 \\
\tilde{\Psi}_L(1+\tfrac{R}{a})=\tilde{\Psi}_{P}(1+\tfrac{R}{a}) && \tilde{\Psi}'_L(1+\tfrac{R}{a})=\tilde{\Psi}'_{P}(1+\tfrac{R}{a}) \nonumber
\end{eqnarray}
where subscript $L$ refers to the region in which the potential is calculated using Laplace's equation, and subscript $P$ refers to the region in which we evaluate the MPB equation. The electric potential is obtained numerically in one single step using eqns (\ref{laplace3}) and (\ref{bclaplace3}). The $\tilde{b}_c$ coefficient and the ionic concentration $n_c(r)$ are determined by eqns (\ref{bc}) and (\ref{nc}), respectively.

\section{Results and discussion}

For all the calculations, the temperature $T$ has been taken equal to 298.15 K and the relative electric permittivity of the suspending liquid $\epsilon_r=$ 78.55, which coincides with that of the deionized water, although no additional ions different to those stemming from the particles have been considered in the present model. Also, the valence of the added counterions $z_c$ has been chosen equal to $+1$ and the particle radius $a=100$ nm. Other values for $z_c$ could have been chosen. The model for point-like ions is able to work with any value of $z_c$ in the Poisson-Boltzmann equation, but we think that the predictions of this model will be less accurate in the case of multivalent counterions, since it is based on a mean-field approach that does not consider ion-ion correlations. Nevertheless, when we take into account the finite size of the ions, the main objective of this work, we include correlations associated with the ionic excluded volume, solving partly this problem, because we are still not considering the electrostatic ion-ion correlations. 

For the sake of simplicity, we assume that  the average volume occupied by a counterion is $V=(2R)^{3}$, being $2R$ the counterion effective diameter. With this consideration, the maximum possible concentration of counterions due to the excluded volume effect is $n_c^{max}=(2R)^{-3}$. This corresponds to a cubic package (52\% packing). In molar concentrations, the values used in the calculations, $n_c^{max}=$ 22, 4 and 1.7 M, correspond approximately to counterion effective diameters of $2R=$ 0.425, 0.75 and 1 nm, respectively. These are typical hydrated ionic radii \cite{Israelachvili1992}. We think also that our model could be useful for short or medium length polyelectrolytes, using an effective value for $n_c^{max}$, as long as different ordered phases do not coexist.

We will discuss the results obtained from three different models, the classical Poisson-Boltzmann equation (PB), the modified Poisson-Boltzmann equation by the finite ion size effect (MPB), and the MPB equation including also the distance of closest approach of the counterions to the surface of the charged particle (MPBL).

\begin{figure}[b!]
\centering
  \includegraphics[height=9.5cm]{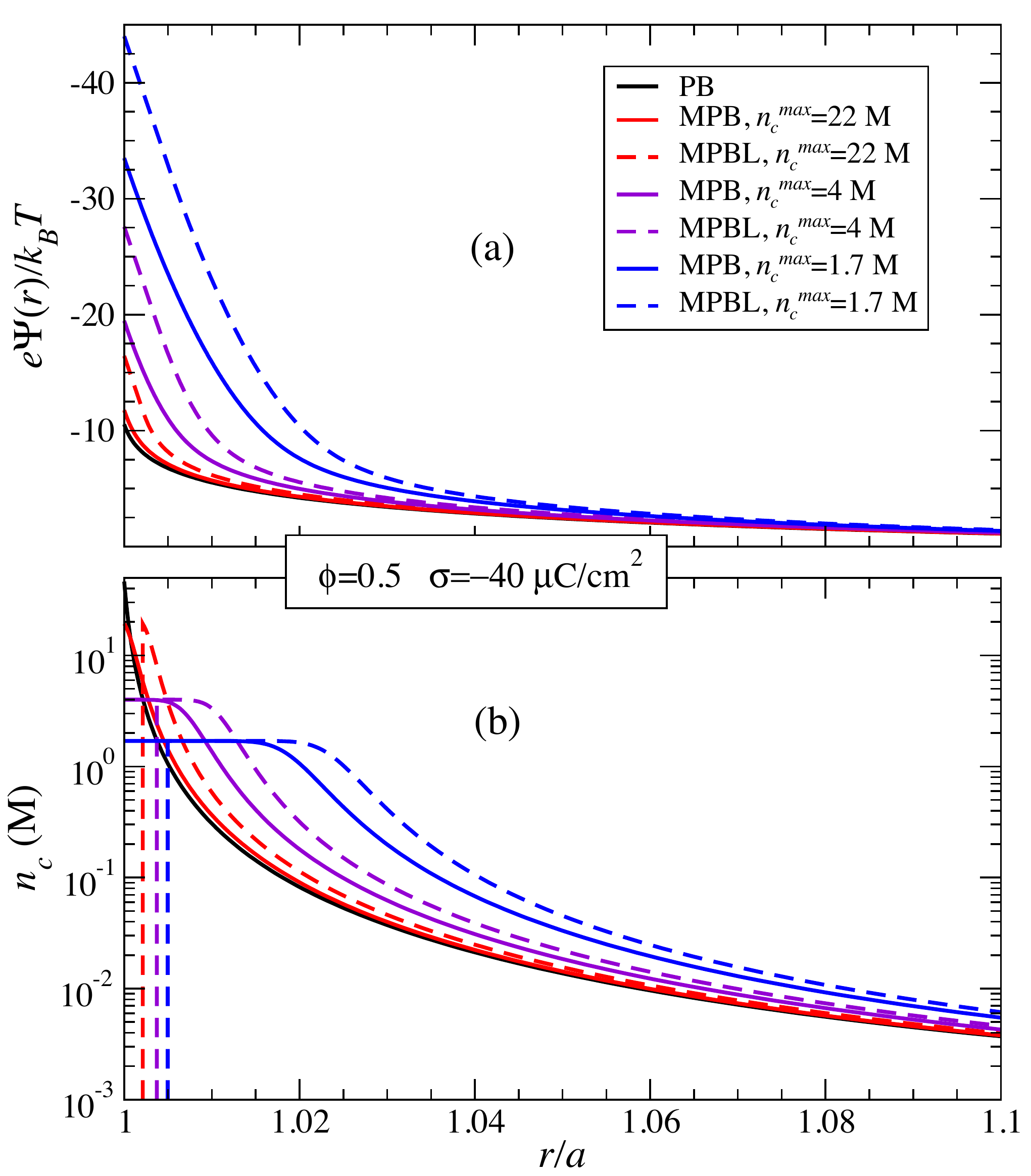}
  \caption{Dimensionless electric potential distribution (a) and counterions concentration (b) along the cell for different ion sizes, considering (dashed lines) or not (solid lines) the excluded region in contact with the particle. Black lines show the results of the PB equation.}
  \label{fgr:sigma40fi005}
\end{figure}

Fig. \ref{fgr:sigma40fi005} shows the dimensionless equilibrium electric potential distribution, Fig. \ref{fgr:sigma40fi005}a, and the counterions concentration profiles, Fig. \ref{fgr:sigma40fi005}b, along the cell. We display in solid black lines the predictions of the PB equation, coloured solid lines stand for the results of the MPB equation and dashed lines correspond to the outcomes of the MPBL model. Different colours stand for different counterion sizes. The particle surface charge density have been chosen equal to $-$40 $\mu$C/cm$^2$ and the particle volume fraction is $\phi=0.5$ (very concentrated suspension), which implies a normalized cell size of $b/a=1.26$.

We can observe in Fig. \ref{fgr:sigma40fi005}a that the inclusion of the finite ion size effect (MPB calculations) always rises the surface electric potential in comparison with the PB predictions, which are recovered in the limit of point-like counterions ($n_c^{max}=\infty$). The reason is that the limitation of the counterions concentration in the vicinity of the particle seriously diminishes the screening of the particle charge, and consequently, there is an increment of the surface potential. An additional significative increase of the surface potential is obtained when we take into account the distance of closest approach of the counterions to the particle surface (MPBL model). As expected, the existence of a Laplace region free of counterions also penalizes the screening of the particle charge.

We can also see in Fig. \ref{fgr:sigma40fi005}b how the finite ion size (MPB) creates saturated regions which corresponds to plateaus in the counterionic concentration profiles. These condensates extend to larger distances when we consider the excluded region in contact with the particle (MPBL). 

\begin{figure}[b!]
\centering
  \includegraphics[height=5.5cm]{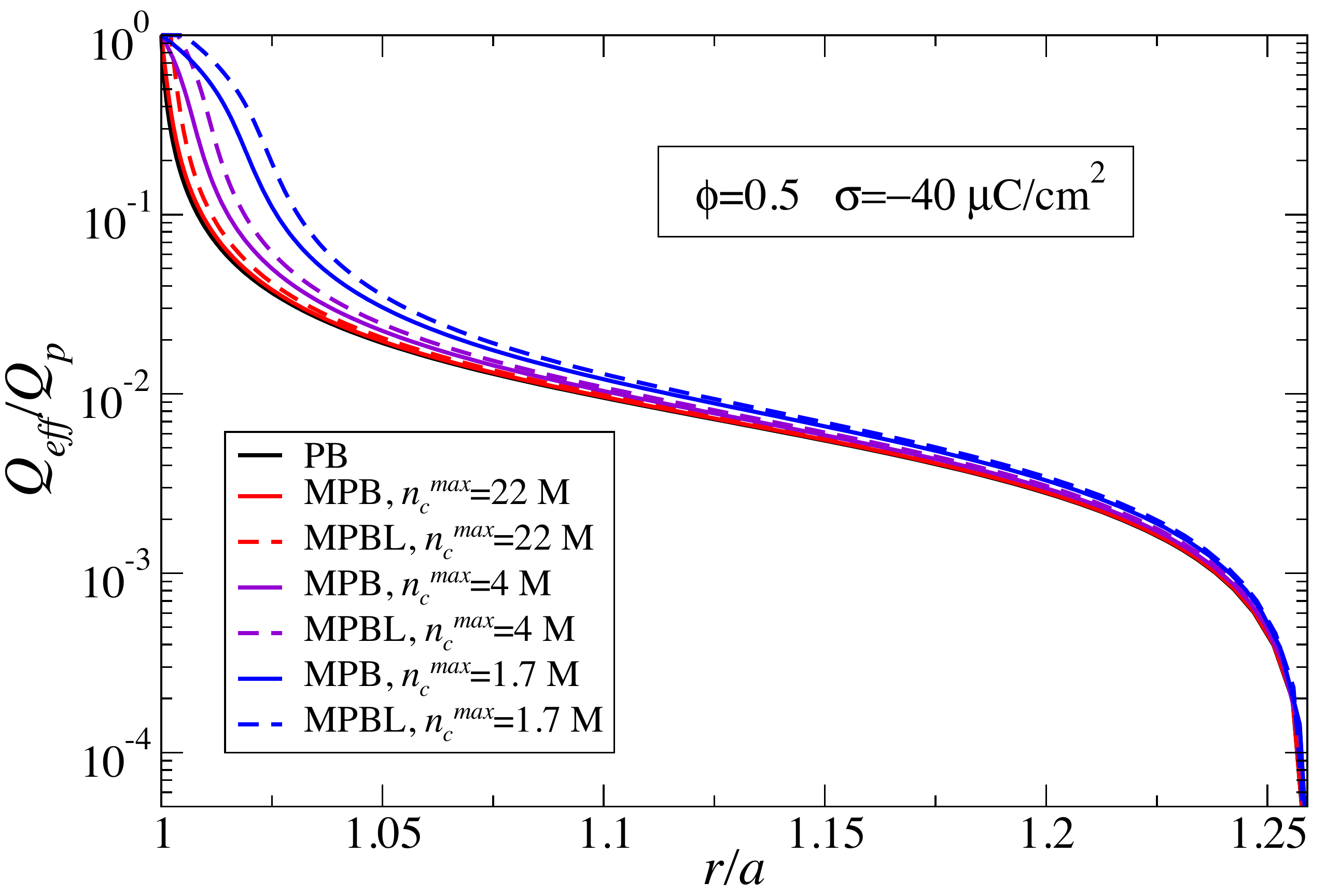}
  \caption{Effective charge divided by the particle charge along the cell for different ion sizes, considering (dashed lines) or not (solid lines) the excluded region in contact with the particle. Black lines show the results of the PB equation.}
  \label{fgr:qef}
\end{figure}

Fig. \ref{fgr:qef} shows the effective charge divided by the particle charge along the cell for a particle volume fraction $\phi=0.5$ and for different ion sizes, accounting or not for the excluded region in contact with the particle. The effective charge is the charge seen at a given radial distance from the center of the particle which includes the own particle charge and the ionic volume charge density in the double layer integrated from the particle surface to $r$ distance \cite{Carrique2009}. From this definition, the ratio $Q_{eff}/Q_p$ is equal to one at the surface of the particle and zero at the outer surface of the cell. The results of this figure reinforced those discussed previously in Fig. \ref{fgr:sigma40fi005}. As expected, the larger the ion size, the softer the screening of the particle charge. Besides, when we consider the distance of closest approach of the counterions to the particle surface, the screening is even less effective at every given distance from the latter. 

\begin{figure}[t!]
\centering
  \includegraphics[height=6.36cm]{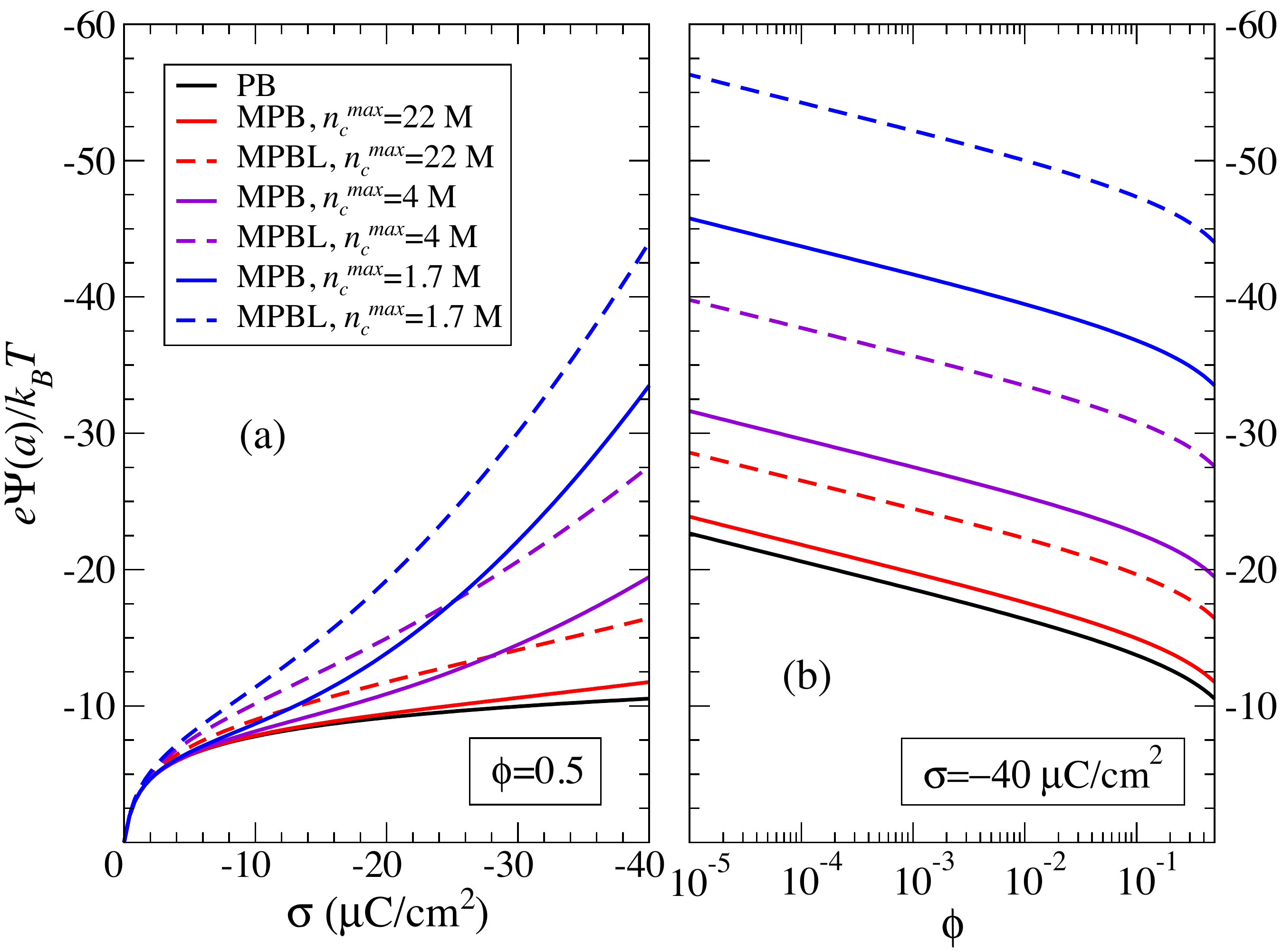}
  \caption{Dimensionless surface electric potential for different values of particle surface charge density (a) and particle volume fraction (b) for different ion sizes, considering (dashed lines) or not (solid lines) the excluded region in contact with the particle. Black lines show the results of the PB equation.}
  \label{fgr:barridossigmafi}
\end{figure}

Fig. \ref{fgr:barridossigmafi} displays the dimensionless equilibrium surface electric potential for a wide range of particle surface charge densities, Fig. \ref{fgr:barridossigmafi}a, and particle volume fractions, Fig. \ref{fgr:barridossigmafi}b. We repeat this study for different counterion sizes. We find that the surface electric potential increases with the particle charge density, Fig. \ref{fgr:barridossigmafi}a. However, in some cases there is a different behaviour in comparison with the point-like case. Initially, a fast and roughly increase of the surface potential with the surface charge density is observed, which is followed by a much slower growth at higher surface charge densities for the PB case or when the size of the counterions is very small. This phenomenon is related to the classical counterion condensation effect: for high surface charges a layer of counterions develops very close to the particle surface \cite{Ohshima2006}. When the ion size is taken into account (MPB) we limit the appearance of the classical condensation effect, because when the surface charge is increased, the additional counterions join the condensate enlarging it. If the region of closest approach of the counterions to the particle surface is also considered (MPBL), the mechanism is the same, but now the additional counterions are located in farther positions from the particle surface. This explains the further increase of the surface potential observed for large ion sizes and high surface charges densities. 

On the other hand, the surface electric potential decreases when the particle volume fraction increases, irrespectively of the case studied: PB, MPB or MPBL, Fig. \ref{fgr:barridossigmafi}b. When the particle concentration raises, the available space for the counterions inside the cell decreases and, consequently, the screening of the particle charge largely augments, thus reducing the value of the surface potential. 

\begin{figure}[t]
\centering
  \includegraphics[height=6.0cm]{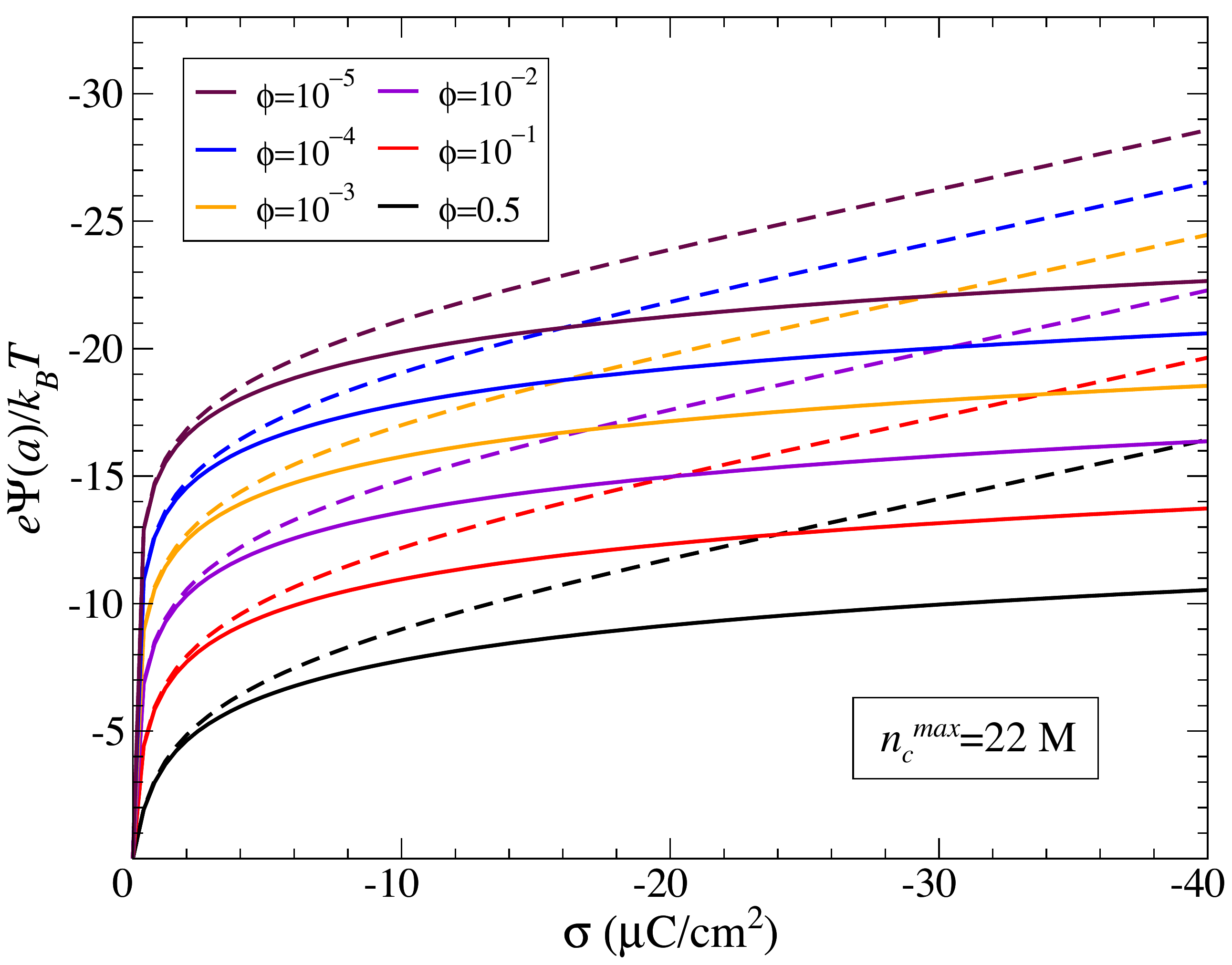}
  \caption{Dimensionless surface potential against the surface charge density for different particle volume fraction values. Solid lines stand for the results of the PB equation. Dashed lines stand for the results of the MPBL model.}
  \label{fgr:barridosigma}
\end{figure}

\begin{figure}[t!]
\centering 
  \includegraphics[height=6.0cm]{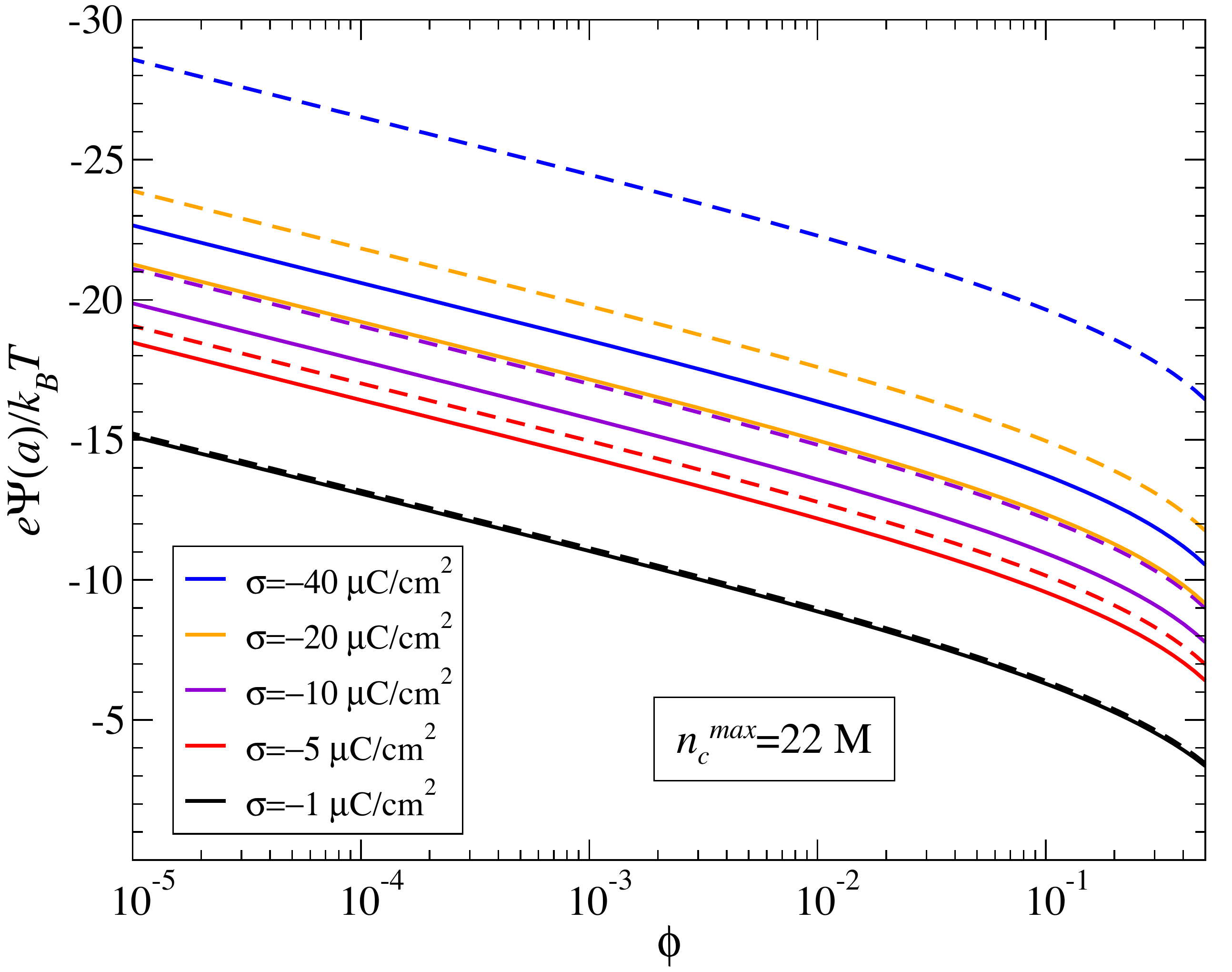}
  \caption{Dimensionless surface potential against the particle volume fraction for different surface charge density values. Solid lines stand for the results of the PB equation. Dashed lines stand for the results of the MPBL model.}
  \label{fgr:barridofi}
\end{figure}

Figs. \ref{fgr:barridosigma} and \ref{fgr:barridofi} expand the results of Fig. \ref{fgr:barridossigmafi}. In them, we compare the results of the PB equation (solid lines) with those of the MPBL model (dashed lines) for a given counterion size (typical of an hydronium ion in solution). Fig. \ref{fgr:barridosigma} presents the dimensionless surface electric potential at a wide range of particle surface charge densities. The different coloured lines correspond to different particle volume fractions. The most remarkable fact shown in Fig. \ref{fgr:barridosigma} is the large influence of the finite ion size effect even for moderately low particle surface charge densities. While for PB predictions the surface potential hardly increases with surface charge for each volume fraction, the MPBL results display an outstanding growth for the same conditions, associated with the lower charge screening ability of the counterions because of their finite size. This fact will surely have important consequences on the electrokinetic properties of such particles in concentrated salt-free suspensions, as has already been shown for dilute suspensions in electrolyte solutions by Aranda-Rasc\'on \textit{et al.} \cite{ArandaRascon2009,ArandaRascon2009b}. 

Finally, Fig. \ref{fgr:barridofi} shows the dimensionless electric potential at the particle surface against the particle volume fraction. The different coloured lines correspond to different negative particle surface charge densities. As previously stated, the particle surface potential monotonously decreases with volume fraction at fixed particle charge density. For the counterion size chosen, the larger the surface charge, the larger the relative increase of the potential at every volume fraction. 

From the results, we think that it is clear that the influence of the finite ion size effect on the EDL description cannot be neglected for many typical particle charges and volume fractions.

\section{Conclusions}
In this work we have studied the influence of finite ion size corrections on the description of the equilibrium electric double layer of a spherical particle in a concentrated salt-free suspension. The resulting model is based on a mean-field approach, due to its reasonable success in modeling electrokinetic and rheological properties of concentrated suspensions, particularly the salt-free ones which deserve a special theoretical interest.

We have used a cell model approach to manage with particle-particle interactions, and modified Poisson-Boltzmann equations (MPB and MPBL) to account for such ion size effects. The theoretical procedure has followed that by Borukhov \cite{Borukhov2004} but with the additional inclusion of a region of closest approach for counterions to the particle surface \cite{ArandaRascon2009,ArandaRascon2009b}. We also have made numerical calculations stressing and discussing the results for concentrated suspensions. The results have clearly shown that the finite ion size effect (MPB equation) is quite important for moderate to high particle charges at every particle volume fraction, and even more if the distance of closest approach of the counterions to the particle surface is taken into account (MPBL model).

This equilibrium model presented in the paper is on the base of future nonequilibrium models of the response of a suspension to external electric fields. Experimental results concerning the DC electrophoretic mobility, dynamic electrophoretic mobility, electrical conductivity and dielectric response, should be compared with the predictions of the latter models to test them. To perform such comparisons highly charged particles are required. Existing highly charged sulfonated polystyrene latexes could be good candidates.

As a final conclusion, we think that it is mandatory to apply these models to improve the existing theories for predicting non-equilibrium properties in concentrated suspensions. This task will be addressed by the authors in the near future.

\appendix

\section*{Appendix: Invariance of MPB equation under changes in the origin of the electric potential}

\renewcommand{\theequation}{A-\arabic{equation}}\setcounter{equation}{0}

In this Appendix we demonstrate that the new MPB equation is invariant under changes in the origin of the electric potential, as it must be according to basic physical grounds. This is not a trivial question, because the classical Poisson-Boltzmann equation for the low particle concentration case in electrolyte solutions is not invariant under changes in the origin of the electric potential: it is only valid when we take the potential zero at an infinite distance from the particle surface (bulk). When we study salt-free and/or concentrated suspensions, we do not have a bulk, because the electroneutrality is achieved at the outer surface of the cell, and we have to take care about the correct invariance of the equation. In this case we do not have an ``infinity'' to set the origin of the potential and we have also an unknown coefficient $b_c$ that must be correctly defined to account for the latter statements. Our PB description for the salt-free case \cite{RuizReina2008} satisfies the above-mentioned invariance requirement.

As we said before, the unknown coefficient $b_c$, that appears in the MPB equation, eqn (\ref{mpb}),  represents the ionic concentration where the equilibrium electric potential is zero. Our choice along this work has been $\Psi(b)=0$. According to this definition and using eqn (\ref{nc}), we can see that $n_c(b)=b_c$. 

\begin{figure}[t]
\centering
  \includegraphics[height=4.0cm]{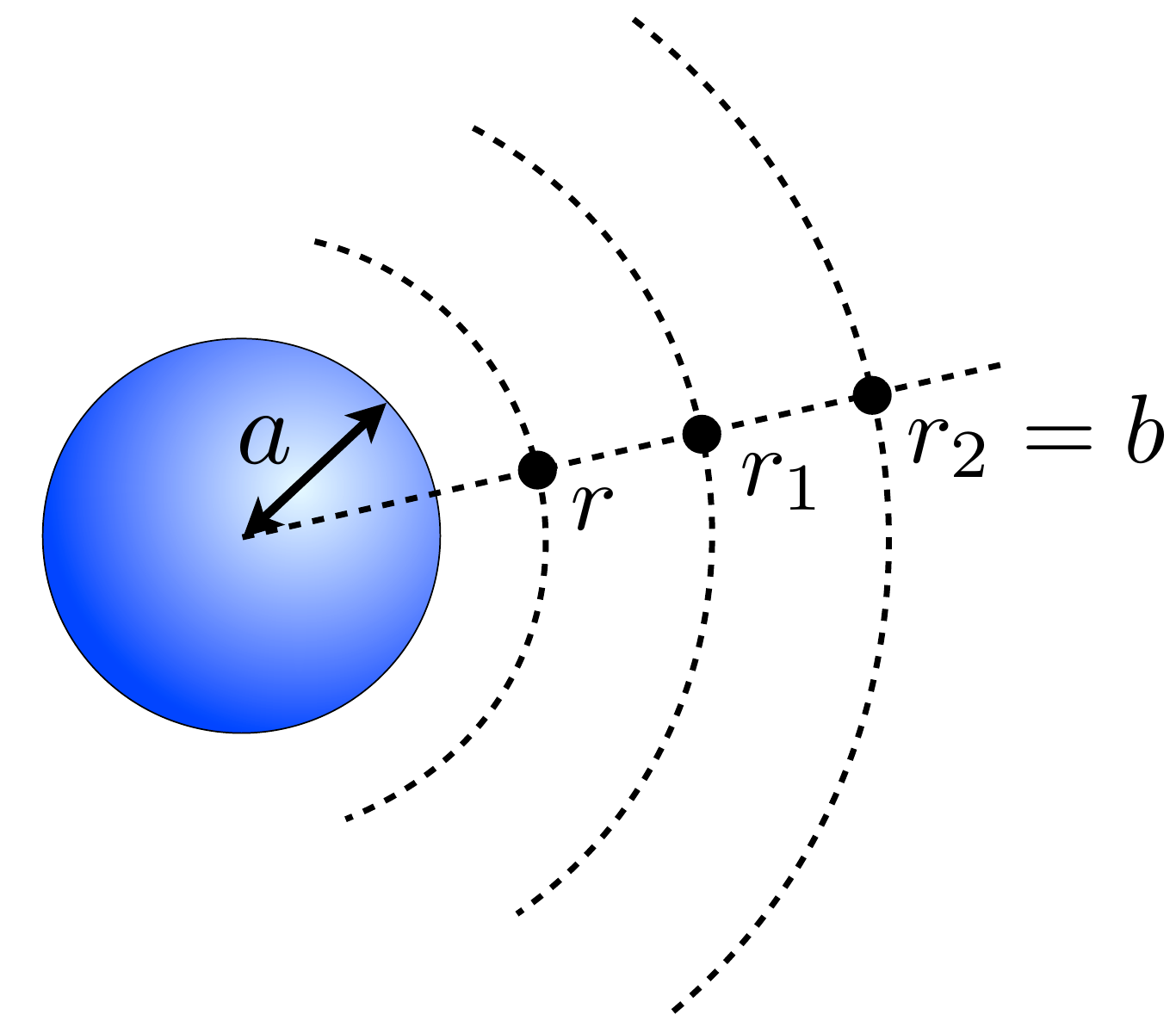}
  \caption{Changes in the origin of the electric potential.}
  \label{fgr:potorigin}
\end{figure}

The potential difference between two points $r$, $r_1$ does not depend on the origin of potentials, Fig. \ref{fgr:potorigin},
\begin{equation}
\Psi(r)-\Psi(r_1)=\Psi^*(r)-\Psi^*(r_1)
\end{equation}
where the magnitudes without asterisks refer to the origin $\Psi(b)=0$ and those with asterisks, to an arbitrary origin, $\Psi^*(r_1)=0$, for the electric potential. The relationship between the potentials referred to both choices of electric potential origins is
\begin{equation}
\Psi(r)=\Psi^*(r)+\Psi(r_1)
\end{equation}

Substituting the last expression in the MPB equation, eqn (\ref{mpb}), we obtain
\begin{align}
&\frac{\mathrm{d}^2\Psi^*(r)}{\mathrm{d}r^2}+\frac{2}{r}\frac{\mathrm{d}\Psi^*(r)}{\mathrm{d}r}\nonumber\\ 
&=-\frac{z_ce}{\epsilon_0\epsilon_r}\frac{b_c\exp\left(-\frac{z_ce\Psi^*(r)}{k_BT}\right)\exp\left(-\frac{z_ce\Psi(r_1)}{k_BT}\right)}{1+\frac{b_c}{n_c^{max}}\left[\exp\left(-\frac{z_ce\Psi^*(r)}{k_BT}\right)\exp\left(-\frac{z_ce\Psi(r_1)}{k_BT}\right)-1\right]}
\end{align}

After a little algebra, and defining
\begin{equation}\label{bc*}
b_c^*=\frac{b_c}{\left(1-\frac{b_c}{b_c^{max}}\right)\exp\left(-\frac{z_ce\Psi(r_1)}{k_BT}\right)+\frac{b_c}{b_c^{max}}}
\end{equation}
it yields
\begin{align}\label{mpb*}
\frac{\mathrm{d}^2\Psi^*(r)}{\mathrm{d}r^2}&+\frac{2}{r}\frac{\mathrm{d}\Psi^*(r)}{\mathrm{d}r}\nonumber\\ 
&=-\frac{z_ce}{\epsilon_0\epsilon_r}\frac{b_c^*\exp\left(-\frac{z_ce\Psi^*(r)}{k_BT}\right)}{1+\frac{b_c^*}{n_c^{max}}\left[\exp\left(-\frac{z_ce\Psi^*(r)}{k_BT}\right)-1\right]}\nonumber\\
&=-\frac{z_ce}{\epsilon_0\epsilon_r}n_c^*(r)
\end{align}
where we have defined
\begin{equation}\label{nc*}
n_c^*(r)=\frac{b_c^*\exp\left(-\frac{z_ce\Psi^*(r)}{k_BT}\right)}{1+\frac{b_c^*}{n_c^{max}}\left[\exp\left(-\frac{z_ce\Psi^*(r)}{k_BT}\right)-1\right]}
\end{equation}

Eqn (\ref{mpb*}) shows that $\Psi^*(r)$ satisfies the MPB equation, eqn (\ref{mpb}), with a different coefficient $b_c^*$. Let's finally check that the concentration of counterions, $n_c^*(r)$, is the same irrespective of the origin of the electric potential: $n_c^*(r)=n_c(r)$, in spite of the different values of the unknown coefficients $b_c$ and $b_c^*$.

By inserting eqn (\ref{bc*}) in eqn (\ref{nc*}), we obtain
\begin{align}
n_c^*(r)=&\frac{b_c}{\left(1-\frac{b_c}{b_c^{max}}\right)\exp\left(-\frac{z_ce\Psi(r_1)}{k_BT}\right)+\frac{b_c}{b_c^{max}}}\nonumber\\
&\cdot\frac{\exp\left(-\frac{z_ce\Psi^*(r)}{k_BT}\right)}{1+\frac{b_c}{n_c^{max}}\displaystyle\frac{\left[\exp\left(-\frac{z_ce\Psi^*(r)}{k_BT}\right)-1\right]}{\left(1-\frac{b_c}{b_c^{max}}\right)\exp\left(-\frac{z_ce\Psi(r_1)}{k_BT}\right)+\frac{b_c}{b_c^{max}}}}\nonumber\\
=&\frac{b_c\exp\left(-\frac{z_ce\Psi(r)}{k_BT}\right)}{1+\frac{b_c}{n_c^{max}}\left[\exp\left(-\frac{z_ce\Psi(r)}{k_BT}\right)-1\right]}=n_c(r)
\end{align}
as we wanted to demonstrate.

\section*{Acknowledgements}
Junta de Andaluc\'ia, Spain (Project P08-FQM-3779), MEC, Spain (Project FIS2007-62737) and MICINN, Spain (Project FIS2010-18972), co-financed with FEDER funds by the EU. Helpful discussions with Dr. Juan J. Alonso are gratefully acknowledged.

\end{document}